\documentclass[aps,prl,reprint,twocolumn,superscriptaddress,longbibliography,nofootinbib,floatfix,showpacs]{revtex4-1}
\usepackage{epsfig}
\usepackage{amsmath,amssymb,amsfonts}
\usepackage{mathrsfs}
\usepackage{bbm}
\usepackage{slashed}
\usepackage{graphicx}
\usepackage{verbatim}

\usepackage[T1]{fontenc}
\usepackage[latin1]{inputenc}
\usepackage[colorlinks]{hyperref}

\usepackage{ifthen}
\usepackage{forloop}
\usepackage[english]{babel}
\usepackage{mathbbol}

\usepackage[usenames]{color}

\usepackage{ulem}

\definecolor{darkgreen}{rgb}{0.2,0.6,0}
\definecolor{lightblue}{rgb}{0,0.5,0.8}
\definecolor{lightred}{rgb}{0.8,0.2,0.2}
\definecolor{darkorange}{rgb}{1,0.549,0}
\newcommand{\be}{\begin{equation}}
\newcommand{\ee}{\end{equation}}
\newcommand{\bw}{\begin{widetext}}
\newcommand{\ew}{\end{widetext}}
\newcommand{\bi}{\begin{itemize}}
\newcommand{\ei}{\end{itemize}}
\newcommand{\bea}{\begin{eqnarray}}
\newcommand{\eea}{\end{eqnarray}}

\newcommand{\cR}{\mathcal R}
\newcommand{\p}{\partial}

\newcommand{\RG}{{\small RG}}

\newcommand{\half}{\tfrac{1}{2}}

\def\cR{{\cal R}}

\newcommand{\Nb}{\bar{N}}
\newcommand{\Nh}{\hat{N}}
\newcommand{\sib}{\bar{\sigma}}
\newcommand{\sh}{\hat{\sigma}}

\newcommand{\Kb}{\bar{K}}

\begin{document}
%-------------------------------------------------------------------

\title{Quantum gravity on foliated spacetime - asymptotically safe and sound}

\author{Jorn Biemans}
\email[]{jbiemans@science.ru.nl}
\affiliation{
Institute for Mathematics, Astrophysics and Particle Physics (IMAPP),\\
Radboud University Nijmegen, Heyendaalseweg 135, 6525 AJ Nijmegen, The Netherlands
}   
\author{Alessia Platania}
\email{alessia.platania@oact.inaf.it}
\affiliation{
	Institute for Mathematics, Astrophysics and Particle Physics (IMAPP),\\
	Radboud University Nijmegen, Heyendaalseweg 135, 6525 AJ Nijmegen, The Netherlands
}
\affiliation{University of Catania, via S. Sofia 63, I-95123 Catania, Italy\\}
\affiliation{INAF, Catania Astrophysical Observatory, via S. Sofia 78, I-95123 Catania, Italy\\}
\affiliation{INFN, Catania Section, via S. Sofia 64, I-95123, Catania, Italy\\}
\author{Frank Saueressig}
\email[]{f.saueressig@science.ru.nl}
\affiliation{
Institute for Mathematics, Astrophysics and Particle Physics (IMAPP),\\
Radboud University Nijmegen, Heyendaalseweg 135, 6525 AJ Nijmegen, The Netherlands
}

\begin{abstract}
Asymptotic Safety provides a mechanism for constructing a consistent and predictive quantum theory of gravity valid on all length scales. Its key ingredient is a non-Gaussian fixed point of the gravitational renormalization group flow which controls the scaling of couplings and correlation functions at high energy. In this work we use a functional renormalization group equation adapted to the ADM-formalism for evaluating the gravitational renormalization group flow on a cosmological Friedmann-Robertson-Walker background. Besides possessing the non-Gaussian fixed points characteristic for Asymptotic Safety the setting exhibits a second family of non-Gaussian fixed points with a positive Newton's constant and real critical exponents. The presence of these new fixed points alters the phase diagram in such a way that all renormalization group trajectories connected to classical general relativity are well-defined on all length scales.  In particular a  positive cosmological constant is dynamically driven to zero in the deep infrared. Moreover, the scaling dimensions associated with the universality classes emerging within the causal setting exhibit qualitative agreement with results found within the $\epsilon$-expansion around two dimensions, Monte Carlo simulations based on Lattice Quantum Gravity, and the discretized Wheeler-deWitt equation.
\end{abstract}
\pacs{}

\maketitle
%%

%--------------------------------------------------------
\section{Introduction}
%--------------------------------------------------------
Obtaining a consistent theory for the gravitational force valid on all length scales is a central challenge in contemporary theoretical high energy physics. 
While general relativity provides a highly successful description of gravity from sub-millimeter up to cosmic scales, it is expected that the theory breaks down for energies around the Planck scale. An interesting mechanism which may provide a short-distance completion of the gravitational force is  Asymptotic Safety \cite{Niedermaier:2006wt,Codello:2008vh,Litim:2011cp,Percacci:2011fr,Reuter:2012id,Reuter:2012xf,Nagy:2012ef}. Building on results for gravity in $2+\epsilon$ spacetime dimensions \cite{Christensen:1978sc,Gastmans:1977ad}, it was first suggested by Weinberg  \cite{Weinberg:1980gg} that the short-distance behavior of gravity may be controlled by a non-Gaussian fixed point (NGFP) of the underlying renormalization group (RG) flow. Provided that such a fixed point exists and comes with a finite number of relevant parameters the resulting
quantum theory of gravity would have the same predictive power as a perturbatively renormalizable quantum field theory. Moreover, 
the proposal  
circumvents the pitfalls encountered in the  perturbative quantization of the Einstein-Hilbert action \cite{'tHooft:1974bx,Goroff:1985sz,vandeVen:1991gw}, which basically originate from the fact that the Gaussian fixed point (GFP) repels the flow of a positive Newton's coupling for increasing energy. As one of its main virtues Asymptotic Safety is conservative in the sense that it stays within the framework of quantum field theory without relying on the introduction of new physics principles beyond the quantum field theory framework. 

Notably, the NGFP underlying the Asymptotic Safety program is not ``put in by hand'': its existence must be established or falsified by explicit computations
of the gravitational RG flow.
% has to be shown from actual computations.
% of the gravitational RG flow. 
 Over the years a series of complementary approaches capable of 
% which are, in principle, suitable for 
 testing the Asymptotic Safety hypothesis have been developed. Starting from the pioneering work \cite{Reuter:1996cp}, functional renormalization group methods have established the existence of a suitable NGFP in a wide range of approximations \cite{Souma:1999at,Lauscher:2001ya,Reuter:2001ag,Litim:2003vp,Fischer:2006fz,Codello:2013fpa,Nagy:2013hka,Becker:2014qya,Gies:2015tca,Donkin:2012ud,Lauscher:2001rz,Lauscher:2002sq,Codello:2006in,Codello:2007bd,Machado:2007ea,Benedetti:2009rx,Benedetti:2009gn,Eichhorn:2009ah,Groh:2010ta,Eichhorn:2010tb,Manrique:2009uh,Manrique:2010mq,Manrique:2010am,Rechenberger:2012pm,Nink:2012vd,Christiansen:2012rx,Christiansen:2014raa,Becker:2014qya,Becker:2014jua,Falls:2014tra,Becker:2014pea,Demmel:2014hla,Christiansen:2015rva,Falls:2015cta}, including the demonstration that the NGFP persists in the presence of the perturbative two-loop counterterm \cite{Gies:2016con} and upon including an infinite number of scale-dependent couplings \cite{Benedetti:2012dx,Demmel:2012ub,Dietz:2012ic,Demmel:2013myx,Dietz:2013sba,Benedetti:2013nya,Demmel:2014sga,Percacci:2015wwa,Borchardt:2015rxa,Demmel:2015oqa,Ohta:2015efa,Ohta:2015fcu,Labus:2016lkh,Henz:2016aoh,Dietz:2016gzg}. Moreover, a first step connecting the NGFP to the underlying conformal field theory appeared in \cite{Nink:2015lmq}, possible completions of the flow at low energy have been discussed in \cite{Reuter:2001ag,Donkin:2012ud,Rechenberger:2012pm,Christiansen:2012rx,Christiansen:2014raa} and geometric arguments determining the scaling of Newton's constant at the NGFP have been forwarded in \cite{Hamber:2004ew,Hamber:2005vc}. In parallel Monte Carlo approaches to quantum gravity including Causal Dynamical Triangulations \cite{Ambjorn:2004qm,Ambjorn:2005db,Ambjorn:2005qt,Ambjorn:2011cg,Anderson:2011bj,Ambjorn:2012jv,Ambjorn:2012ij,Ambjorn:2014gsa,Ambjorn:2016cpa}, Euclidean Dynamical Triangulations \cite{Laiho:2011ya,Smit:2013wua,Ambjorn:2013eha,Rindlisbacher:2015ewa,Laiho:2016nlp} and Lattice Quantum Gravity \cite{Hamber:1992df,Hamber:2015jja} made vast progress towards constructing phase diagrams at the non-perturbative level. While it is conceivable that all of these approaches probe the same universal short distance physics, a unified picture has yet to emerge.

In this work we report significant progress towards both developing a unified picture from these distinguished programs and constructing complete RG trajectories relevant for describing our world. Building on the functional renormalization group approach \cite{Reuter:1996cp}, we study gravitational RG flows in the Arnowitt-Deser-Misner (ADM) variables \cite{Arnowitt:1959ah,Arnowitt:1962hi}. The 
 ADM-formalism imprints spacetime with a foliation structure.
 The resulting distinguished time direction allows the continuation of the flow equation from Euclidean to Lorentzian signature metrics. The phase diagram resulting from projecting the resulting flow onto the (Euclidean) Einstein-Hilbert action evaluated on a Friedmann-Robertson-Walker background
  exhibits a  remarkable combination of highly desirable features. Firstly, the flow possesses a NGFP suitable for Asymptotic Safety. For dimensions $D\le 3.25$ it also furnishes a dynamical mechanism providing the low-energy completion of the renormalization group trajectories with a positive cosmological constant which are of central relevance for describing our world \cite{Reuter:2004nx}. Moreover, the universality classes of the system exhibit universal scaling behaviors which are in qualitative agreement with 
  other approaches to quantum gravity. These findings support the idea \cite{Loll:2000my} that a causal structure may play an important role for obtaining a well-defined long-distance completion of gravity from the quantum theory. 

%--------------------------------------------------------
\section{Functional renormalization}
%--------------------------------------------------------
The functional RG provides a powerful tool for investigating the emergence of critical behavior and the phase structure of a physical system. In its formulation based on the effective average action $\Gamma_k$  \cite{Wetterich:1992yh}, the functional RG equation (FRGE)
\be\label{FRGE}
k\p_k \Gamma_k = \tfrac{1}{2} \, {\rm Str} \left[ \left( \Gamma_k^{(2)} + \cR_k \right)^{-1} \, k \p_k \cR_k \right] \, , 
\ee
implements Wilson's idea that the RG flow of a theory originates from integrating out quantum fluctuations shell-by-shell in momentum space. The interplay of the scale-dependent regulator $\cR_k$ in the numerator and denominator thereby ensures that the change of $\Gamma_k$ is driven by fluctuations with momenta close to $k$. The two-point correlator $\Gamma_k^{(2)}$ makes \eqref{FRGE} a formally exact equation with the same information content as the path integral from which it is derived. A central advantage of the FRGE is that it allows for approximations which do not rely on the existence of a small expansion parameter. Moreover, RG flows can be constructed without the need of specifying a fundamental action. This makes the framework ideally suited for investigating fixed points of the flow which encode the universal scaling behavior of the system.

In the present investigation we use the formulation of the FRGE where the gravitational degrees of freedom are carried by the ADM-fields \cite{Manrique:2011jc,Rechenberger:2012dt}. 
In this case, the (Euclidean) spacetime metric is decomposed according to
\be\label{fol1}
ds^2 
=  N^2 dt^2 +  \sigma_{ij} \, (dx^i + N^i d t) 
(dx^j +  N^j d t)  
\ee 
with the Lapse function $N(t,\vec x)$, the shift vector $N^i(t,\vec x)$ and a metric on the spatial slices $\sigma_{ij}(t,\vec x)$ being functions of the (Euclidean) time $t$ and the spatial coordinates $\vec{x}$. The flow equation is then constructed via the background field method, splitting the component fields into a fixed background and fluctuations
\be
N = \bar{N} + \hat{N} \, , \; \; 
N^i = \bar{N}^i + \hat{N}^i \, , \; \;
\sigma_{ij} = \sib_{ij} + \hat{\sigma}_{ij} \, .  
\ee
The resulting construction is then invariant under foliation-preserving diffeomorphisms, corresponding to a subgroup of the full diffeomorphism group.

%--------------------------------------------------------
\section{Projection of the flow equation}
%--------------------------------------------------------
We study the gravitational RG flow projected on the Einstein-Hilbert action written in terms of the extrinsic curvature $K_{ij} \equiv (2N)^{-1}\left(\p_t \sigma_{ij} - D_i N_j - D_j N_i \right)$, $K \equiv \sigma^{ij} K_{ij}$, and the intrinsic curvature on the $d$-dimensional spatial slices $R$
\be\label{EHaction}
\Gamma_k^{\rm EH} = \tfrac{1}{16 \pi G_k} \! \int \! \!dt d^dx N \sqrt{\sigma} \left[K_{ij} K^{ij} - K^2 - R + 2 \Lambda_k \right] . 
\ee
The gravitational sector is supplemented by a novel gauge-fixing term
\be\label{gf:ansatz}
\Gamma_k^{\rm gf} = \frac{1}{32 \pi G_k} \int dt d^dx \, \sqrt{\sib} \, \left[ F_i \, \sib^{ij} F_j + F^2 \right] \, . 
\ee
The functionals $F$ and $F_i$ are linear in the fluctuation fields,
\be\label{Ggf}
\begin{split}
	F = & \, \p_t \, \Nh +  \p^i \Nh_i - \half \p_t \sh  + \tfrac{2(d-1)}{d}  \Kb \Nh , \\
	F_i = & \, \p_t  \Nh_i - \p_i  \Nh - \half \p_i  \sh + \p^j  \sh_{ji} + (d-2) \Kb_{ij} \Nh^j ,
\end{split}
\ee
where $\hat{\sigma} \equiv \bar{\sigma}^{ij} \hat{\sigma}_{ij}$. The
relations \eqref{Ggf} can be understood as first order differential equations fixing the fluctuations of the Lapse function and shift vector in terms of the other fluctuation fields and the background. As its main virtue the gauge choice \eqref{Ggf} equips all component fields including the Lapse function, the shift vector, and the Faddeev-Popov ghost fields with regular propagators.\footnote{For a related construction in the context of Lorentz-violating field theories see \cite{Anselmi:2008bq,Barvinsky:2015kil}.} This condition actually fixes the gauge uniquely. On a flat Euclidean background all component fields obtain an identical dispersion relation such that, upon Wick-rotation, all fields propagate with the same speed of light.  The ansatz for $\Gamma_k$ is completed by the standard ghost term exponentiating the Faddeev-Popov determinant arising from \eqref{gf:ansatz}. Notably, this is the first time that a completely regular off-shell flow equation based on the ADM-formalism is obtained. 

The evaluation of the trace utilizes the technology of the universal
\RG{} machine \cite{Benedetti:2010nr} combined with the heat-kernel methods on foliated spacetimes \cite{Nesterov:2010yi,D'Odorico:2015yaa} and the off-diagonal heat-kernel technology \cite{Barvinsky:1985an,Decanini:2005gt,Anselmi:2007eq,Groh:2011vn,Groh:2011dw}.  The evaluation is simplified by working on a flat Friedmann-Robertson-Walker background
\be
\Nb = 1 \, , \; \; \Nb_i = 0 \, , \; \; \sib_{ij} = a^2(t) \, \delta_{ij} \, , 
\ee
and parameterizing the fluctuations in terms of the field-decomposition used in cosmic perturbation theory. The flow of Newton's coupling and the cosmological constant is then read off from the coefficient multiplying the extrinsic curvature and volume term, respectively.

%------------------------------------------------------------------------
\section{\texorpdfstring{$\beta$}{beta} functions}
%------------------------------------------------------------------------
The RG flow resulting from the ansatz \eqref{EHaction} is conveniently expressed
in terms of the dimensionless couplings $g_i \equiv \{\lambda,g\}$ and the anomalous dimension $\eta_N$ of Newton's coupling
\be\label{dimless}
\lambda_k \equiv \Lambda_k \, k^{-2} \, , \; g_k \equiv G_k \, k^{d-1} \, , \; \eta_N \equiv (G_k)^{-1} \, k\p_k G_k \, , 
\ee
and encoded in the $\beta$ functions
\be
k \p_k g_i \equiv \beta_{g_i}(\lambda,g) \, . 
\ee
The details of the derivation will be presented elsewhere \cite{Biemans:inprep}.
For a Type I cutoff \cite{Codello:2008vh}, dressing the Laplacian according to $\Delta \mapsto \Delta + R_k$, and choosing $R_k$ as the Litim regulator \cite{Litim:2001up}, the $\beta$ functions for a $d$-dimensional spatial slice read
\be\label{betafunction}
\begin{split}
	\beta_g = & \, (d-1+\eta_N) \, g \, , \\
	\beta_\lambda = & \, (\eta_N - 2) \lambda - \tfrac{g}{(4 \pi)^{(d-1)/2}} \Big[   \tfrac{8}{\Gamma((d+1)/2)} \\ & \;
	- 
	\left( d + 
	\tfrac{d^2 + d -4}{2(1-2\lambda)}
	+ \tfrac{3d-3 - (4 d-2) \lambda }{B_{\rm det}(\lambda)}
	\right) \times \\ & \;
	\left( \tfrac{2}{\Gamma((d+3)/2)} - \tfrac{\eta_N}{\Gamma((d+5)/2)} \right)   \Big] \, , 
\end{split}
\ee
where $B_{\rm det}(\lambda) \equiv  (1 - 2 \lambda) (d - 1 - d \lambda)$.
The explicit form of $\eta_N$ is given by
\be\label{etaflow}
\begin{split}
	\eta_N = \frac{16 \pi g \, B_1(\lambda)}{(4\pi)^{(d+1)/2} + 16 \pi g \, B_2(\lambda)}
\end{split}
\ee
where
\be
\begin{split}
	& B_1(\lambda)   \equiv 
	- \tfrac{d^4 + 14 d^3 - d^2 + 94 d + 12}{12 \, d ( d - 1) \, \Gamma(( d+3)/2)}
	+ \tfrac{d^2+d-4}{12 \, (1-2\lambda) \, \Gamma((d+1)/2)} \\ & \,
	- \tfrac{d^4-15d^2+28d-10}{2 d (d-1) \, (1-2\lambda)^2 \, \Gamma((d+3)/2)}
	%
	%\\ & \,
	+ \tfrac{3d-3 - (4d-2) \lambda }{6 \, B_{\rm det}(\lambda) \,  \Gamma(( d+1)/2)} \\ & \,
	+ \tfrac{c_{1,0} + c_{1,1} \lambda + c_{1,2} \lambda^2 + c_{1,3} \lambda^3 + c_{1,4} \lambda^4 }{4 \, d \,  ( d^2+2d-3) \, B_{\rm det}(\lambda)^2 \,  \Gamma((d+3)/2)}
\end{split}
\ee
and
\be
\begin{split}
 & 	B_2(\lambda) =  \, 
	\tfrac{d^3 - 9 d^2 + 12d+12}{24 \, d \, \Gamma((d+5)/2)}
	+ \tfrac{d^2+d-4}{24 \, (1-2\lambda) \, \Gamma((d+3)/2)} \\ & \, 
	- \tfrac{d^4-15d^2+28d-10}{4 \,d(d-1) \, (1-2\lambda)^2 \, \Gamma((d+5)/2)} 
	 + \tfrac{3d-3 - (4d-2)\lambda}{12 \, B_{\rm det}(\lambda) \, \Gamma((d+3)/2)}  \\ & 
	+ \tfrac{c_{2,0} + c_{2,1} \lambda + c_{2,2} \lambda^2 }{8 \, d  \, B_{\rm det}(\lambda)^2 \,  \Gamma((d+5)/2)} \, . 
\end{split}
\ee
The coefficients $c_{i,j}$ are polynomials in $d$ and read
\be
\begin{split}
c_{1,0} = & \,-(d-1) (5 d^4 - 7 d^3 - 74 d^2 + 56 d -16 ) \, ,  \\
c_{1,1} = & \, 4 (d-1) (d^4 - 7 d^3 - 62 d^2+ 16 d -16)
 \, ,  \\
c_{1,2} = & \, 4 d^5 + 32 d^4 + 388 d^3 - 232 d^2 -64 d -64  \, ,  \\
c_{1,3}	= & \,-128 \, d (d+1) (3 d-2)   \, , \\
c_{1,4}	= & \, 128 \, d^2 \, (d+1)   \, .		
\end{split}
\ee
and
\be
\begin{split}
	c_{2,0} = & \, - 5 d^3 + 22 d^2 -24 d + 16      \, ,  \\
	c_{2,1} = & \, 4 d^3 - 40 d^2 + 64 d - 64    \, ,  \\
	c_{2,2} = & \, 4 d^3  +24 d^2 - 64 d + 64  \, . 
\end{split}
\ee
The $\beta$ functions \eqref{betafunction} together with the anomalous dimension of Newton's constant \eqref{etaflow} completely encode the RG flow resulting from the ansatz \eqref{EHaction}.

%------------------------------------------------------------------------
\section{Fixed points and universality classes}
%------------------------------------------------------------------------
The RG encodes universal critical behavior of a system in the fixed points $g_{i,*}$ of the $\beta$ functions, $\beta_{g_i}|_{g_{i,*}} = 0$. The flow in the vicinity of a fixed point is governed by the stability matrix
$B_{ij} \equiv \partial_{g_j} \beta_{g_i}|_{g_*}$ resulting from 
 linearizing the beta functions at $g_{i,*}$. The stability coefficients $\theta_i$, defined as minus the eigenvalues of $B_{ij}$, indicate whether 
 the corresponding eigendirection of the fixed point attracts (Re  $\theta_i >0$) or repels (Re $\theta_i < 0$) the flow for increasing energy. UV-attractive directions are associated with relevant parameters of the theory which need to be fixed by experiment. Moreover, the stability coefficients contain characteristic information about the underlying universality class which allows a direct comparison based on different computational approaches.

Remarkably, the $\beta$ functions \eqref{betafunction} give rise to a rich fixed point structure which varies with the spacetime dimension $D=d+1$. Firstly, there is the Gaussian Fixed Point (GFP) located in the origin and characterized by classical scaling dimensions. As a consequence the GFP is a saddle-point: trajectories with a positive Newton's coupling are not captured by the GFP at high energy. This reflects the perturbative non-renormalizability of the Einstein-Hilbert action in the Wilsonian language.

\begin{figure}[t]
	\includegraphics[width=0.46\textwidth]{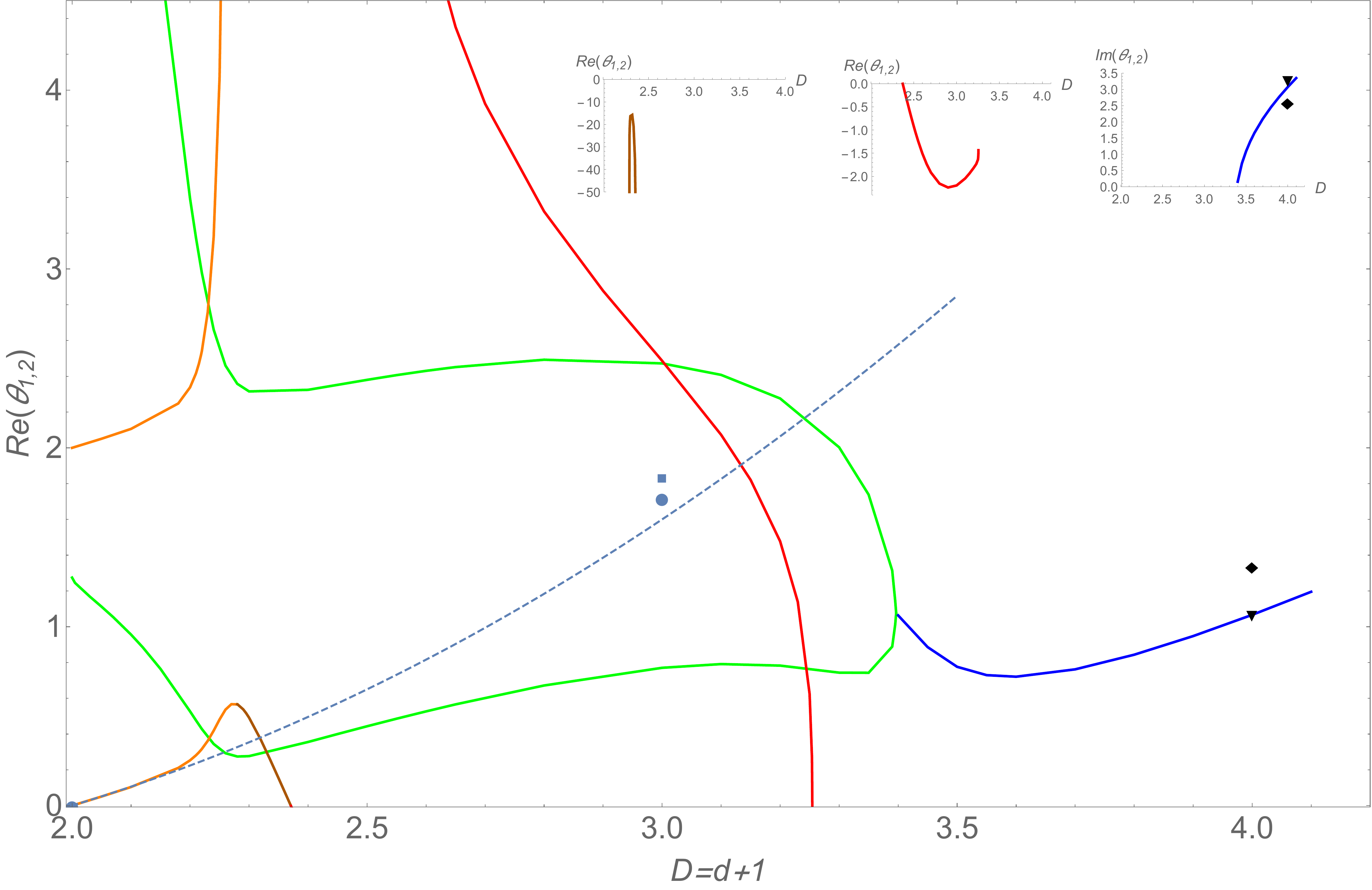}
	\caption{Stability coefficients of the two families of fixed points emerging from the $\beta$ functions \eqref{betafunction} as a function of the spacetime dimension $D=d+1$. The dashed line gives the result from the two-loop epsilon expansion \cite{Aida:1996zn}, the circles indicate the scaling of Newton's coupling found within lattice Quantum Gravity \cite{Hamber:1992df} and the square marks the scaling found from the exact solution of the discretized Wheeler-deWitt equation \cite{Hamber:2011cn,Hamber:2012zm}. In $D=4$ the down-triangle indicates the critical exponents obtained from foliated spacetimes using the Matsubara formalism \cite{Manrique:2011jc} while the diamond corresponds to the dynamical fixed point seen in the ``geometrical'' flow equation \cite{Donkin:2012ud}.}
	\label{fig:3}
\end{figure}

In addition, there are two families of Non-Gaussian Fixed Points (NGFPs) whose most important properties are summarized in Fig.\ \ref{fig:3}.
In $D=2+\epsilon$ dimensions a NGFP with two real, positive stability coefficients (orange line) emerges from the GFP. Its critical exponents agree with the epsilon-expansion of perturbative gravity around two dimensions \cite{Aida:1996zn} to leading order. This universality class has an upper critical dimension $D = 2.28$. At this point there is a transition to a family of saddle points (SP) characterized by a small UV-attractive and a large UV-repulsive critical exponent (brown line). In $D= 2.37$ these critical exponents swap sign, giving rise to the red line of SP-NGFPs existing for $2.37 \le D \le 3.25$. Simultaneously, there is a second family of fixed points (green line) with two real, positive critical exponents. At $D=3.40$ the two stability coefficients coincide at $\theta_1 = \theta_2 = 1.08$. For $D > 3.40$ the real stability coefficients become complex (blue line) which reflects the typical characteristics of the UV-NGFP seen in the functional RG approach to Asymptotic Safety. This universality class can be traced up to $D \approx 40$, where $g_*$ becomes exponentially large. The additional information displayed in Fig.\ \ref{fig:3} encodes the stability coefficients found within related quantum gravity programs. The qualitative agreement of the scaling behavior seen within discrete (Monte Carlo) methods in $D=2+1$ dimensions and the continuum RG is an important indicator for the robustness of the underlying universality classes.

The properties of the NGFPs in $D=2+1$ and $D=3+1$ are summarized as follows. In $D=2+1$, the UV-NGFP and SP-NGFPs are located at
\be
\begin{split}
\mbox{UV-NGFP:} \; \;  & g_* = 0.16, \, \; \lambda_* = 0.03, \, \; g_* \lambda_* = 0.005 , \\
\mbox{SP-NGFP:} \; \; & 
g_* = 0.32, \, \; \lambda_* = 0.20, \, \; g_* \lambda_* =0.07,	
\end{split} 
\ee
and come with stability coefficients
\be\label{uvfp}
\begin{split}
\mbox{UV-NGFP:} & \; \; \; \theta_{1} = 2.47  \, , \quad  \theta_2 = 0.77 \, ,  \\
\mbox{SP-NGFP:} & \; \; \; \theta_{1} = 2.49  \, , \quad \theta_2 = -2.20 \, . 
\end{split}
\ee
In $D=3+1$ there is a unique NGFP for positive Newton's constant at
\be
\begin{split}
	\mbox{AS-NGFP:} \; \;  & g_* = 0.90, \, \; \lambda_* = 0.24, \, \; g_* \lambda_* = 0.21 \, , 	
\end{split} 
\ee
and coming with critical exponents
\be
\mbox{AS-NGFP:} \; \; \theta_{1,2} = 1.06 \pm 3.07 i \, .  
\ee
This NGFP exhibits
the typical complex pair of critical exponents familiar  
 from evaluating the 
flow equation in the metric formulation \cite{Souma:1999at,Lauscher:2001ya,Reuter:2001ag,Litim:2003vp,Fischer:2006fz,Donkin:2012ud,Codello:2013fpa,Nagy:2013hka,Becker:2014qya,Gies:2015tca}. In particular, there is a very good agreement with the critical exponents obtained for foliated spacetime via the Matsubara formalism \cite{Manrique:2011jc,Rechenberger:2012dt}.
Thus it is highly conceivable that the AS-NGFP seen in these computations is the one underlying Asymptotic Safety. 
%

%------------------------------------------------------------------------
\section{Phase diagrams}
\begin{figure}
\includegraphics[width=0.49\textwidth]{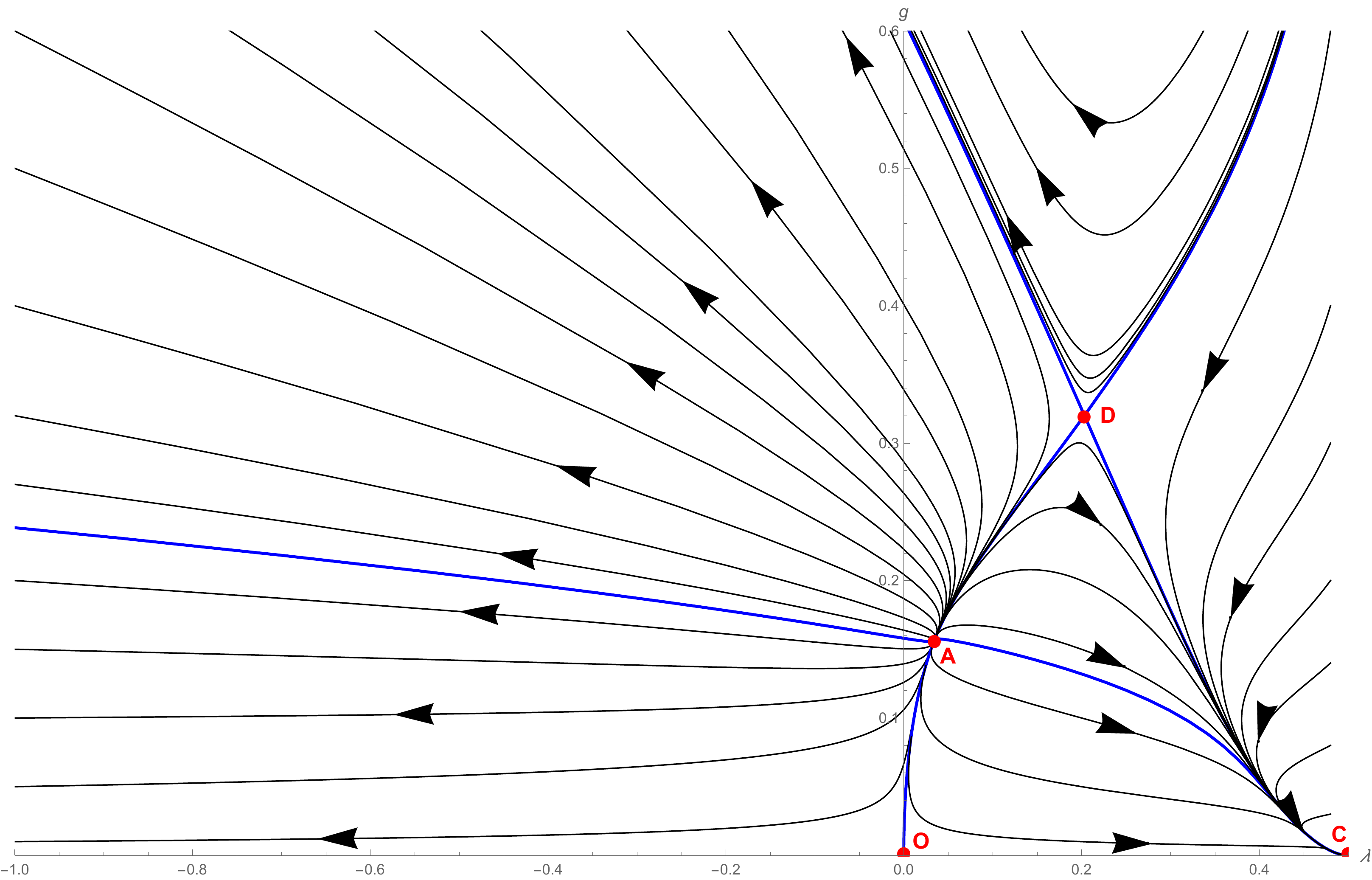} \\[2ex]
\includegraphics[width=0.49\textwidth]{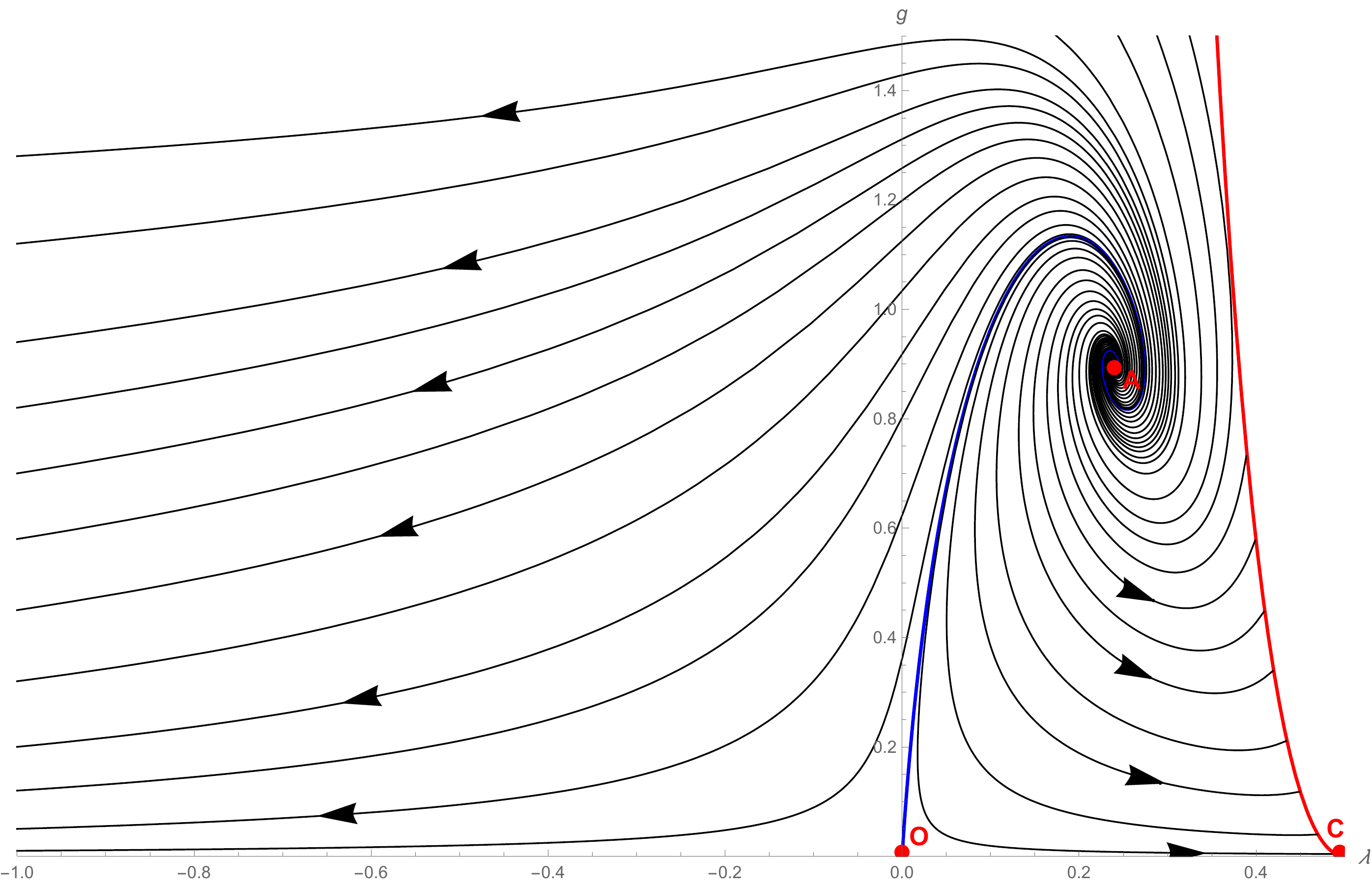}
\caption{Phase diagrams resulting from the $\beta$ functions \eqref{betafunction} for $D=2+1$ (top) and $D=3+1$ (bottom) spacetime dimensions. The GFP, NGFP and SP-NGFP are marked by the points ``O'', ``A'', and ``D''. In $D=2+1$ the interplay of the fixed points ``A'' and ``D'' ensures that the  QFP ``C'' provides the long-distance completion of the renormalization group trajectories with a positive cosmological constant. In $D=3+1$ the SP-NGFP ``D'' is absent and the corresponding trajectories terminate in a divergence of $\eta_N$ (red line).}
\label{fig:1}
\end{figure}

%------------------------------------------------------------------------
%
The phase diagram resulting from the numerical integration of the $\beta$ functions \eqref{betafunction} in $D=2+1$ and $D=3+1$  is shown in the top and bottom panel of Fig.\ \ref{fig:1}.
For $D=2+1$ dimensions, the flow is governed by the interplay of the GFP, the two NGFPs and the two quasi fixed points (QFPs) ``B'' and ``C'' located at $(\lambda,g) = (-\infty,0)$ and $(\lambda,g) = (1/2,0)$, respectively. The phase transition lines connecting these points are depicted in blue. All RG trajectories located below the transition line $\overline{CDB}$ are complete: they are well-behaved for all values $k \in [0,\infty]$. Their high-energy behavior is controlled by the UV-NGFP ``A''. Lowering the RG scale they cross over to the GFP or the saddle point ``D''. Trajectories passing sufficiently close to the GFP develop a classical regime where the Newton coupling and cosmological constant are independent of the RG scale. Depending on whether the flow approaches ``B'', ``O'', or ``C'' the classical regime exhibits a negative, zero, or positive cosmological constant.

The picture found in $D=3+1$ dimensions is similar: the UV completion of the flow is controlled by the AS-NGFP and the classical regime emerges from the crossover of the flow to the GFP. As a consequence of the missing SP-NGFP ``D'' the trajectory $\overline{CD}$ in $D=2+1$ is replaced by a line of singularities where $\eta_N$ diverges (red line). In this case solutions with a positive cosmological constant terminate at a finite RG scale.

Since the RG trajectory describing our world should exhibit a classical regime with a positive cosmological constant \cite{Reuter:2004nx}, it is worthwhile to investigate the mechanism providing the IR completion of these trajectories in  detail. 
\begin{figure}
	\includegraphics[width=0.49\textwidth]{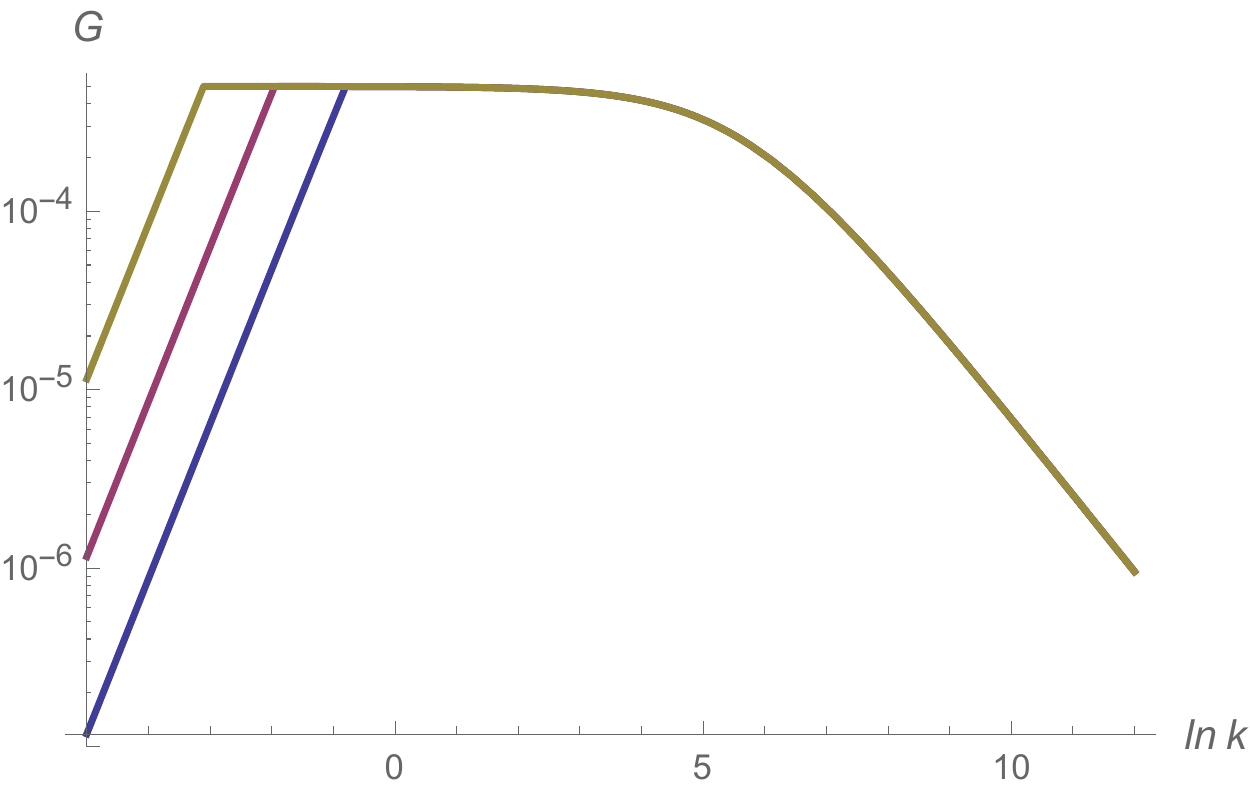} \\[2ex]
	\includegraphics[width=0.49\textwidth]{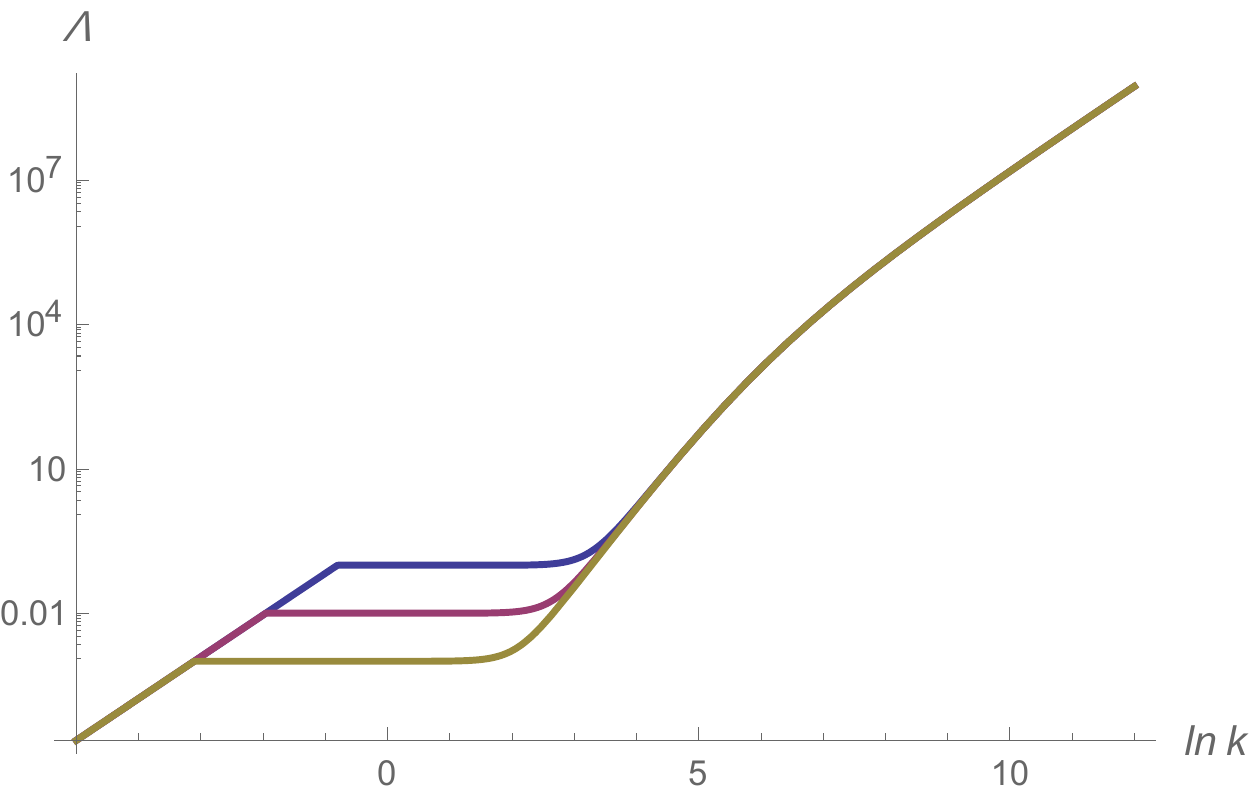}
	\caption{Scale dependence of Newton's constant $G_k$ (top) and the cosmological constant (bottom) for a set of RG trajectories exhibiting a classical regime with a positive cosmological constant for $D=2+1$. The QFP ``C'' drives the infrared-value of $\Lambda$ to zero dynamically.  }
	\label{fig:2}
\end{figure}
Fig.\ \ref{fig:2} displays the typical scale-dependence of the Newton coupling (top) and the cosmological constant (bottom) for this class of solutions. Starting from the high-energy part flowing towards low energy the trajectories undergo four phases: the fixed point regime is controlled by the UV-NGFP ensuring the absence of unphysical UV divergences. Subsequently, there is a semi-classical regime followed by a classical regime where the Newton constant and cosmological constant exhibit plateaus. These features are independent of the spacetime dimension. In $D=2+1$ the low-energy completion of the solutions is provided by a novel phase where both $G$ and $\Lambda$ are dynamically driven to zero. This phase is controlled by the QFP ``C'' situated at $(\lambda,g) = (1/2,0)$. At this point the $\beta$ functions \eqref{betafunction} are ambiguous owed to terms of the form $g/(1-2\lambda)^2$ where both the numerator and denominator vanish. 
RG trajectories approaching ``C'' resolve this ambiguity in such a way that $\lim_{k \rightarrow 0} g/(1-2\lambda)^2 = 5 \pi/6$. In this way ``C'' is turned into an low-energy attractor where $\eta_N = 2$. The mechanism providing the low-energy completion in this sector is essentially the same as the one reported in \cite{Christiansen:2012rx}. In the present case, regularity appears at the level of the background couplings and a Type I regulator though.

Fig.\ \ref{fig:2} also illustrates that the value of the cosmological constant in the classical regime is a free parameter and may be set to its experimentally determined value. The transition (resp.\ termination) scale $k_{\rm t}$ from the classical to the low-energy phase can be estimated from the relation $\Lambda = \lambda k^2$. Setting $\lambda = 1/2$ and substituting the measured value of the cosmological constant $\Lambda_{\rm obs} = 1.19 \times 10^{-52}$ m$^2$ yields $k_t = 3 \times 10^{-33}$ eV, corresponding to a length scale of $l_{\rm t} = 6 \times 10^{25}$ m.  This value is within the radius of the observable universe opening the possibility that effects of a new gravitational phase, possibly related to the physics of a QFP similar to ``C'', may be visible on cosmic scales.

%------------------------------------------------------------------------
\section{Conclusions}
%------------------------------------------------------------------------
In this work we provided a novel, completely regular off-shell formulation of the functional renormalization group equation \eqref{FRGE} on spacetimes carrying a foliation structure. The framework implements the gravitational degrees of freedom in terms of ADM-variables and is ideally suited for computing real-time correlation functions, e.g., in a cosmological context.
As a first application we studied the scale-dependence of Newton's coupling and the cosmological constant in the presence of a foliation structure. The resulting fixed point structure and universal critical exponents
match the leading correction obtained from the perturbative computations in $2+\epsilon$ spacetime dimensions and show a good agreement with lattice quantum gravity in $D=2+1$ dimensions \cite{Hamber:1992df} and the discretized Wheeler-de Witt equation \cite{Hamber:2011cn,Hamber:2012zm}. In particular the real critical exponents of the non-Gaussian fixed points in $D=2+1$ may provide 
a natural explanation for the apparent mismatch between the real critical exponents seen in Monte Carlo approaches and the complex critical exponents typically obtained from the functional renormalization group. A particular interesting feature of the gravitational flow appears in $D=2+1$ dimensions where the interplay of two non-Gaussian fixed points actually resolves the IR singularities typically encountered in the metric formulation. Based on this mechanism \textit{all} renormalization group trajectories exhibiting a classical regime with a positive cosmological constant are well-defined on all length-scales. Presupposing that a similar mechanism is also operative in $D=3+1$ dimension, the framework predicts a distinct modification of general relativity at cosmic distances.
\bigskip

\acknowledgments

We thank D.\ Becker, A.\ Bonanno, G.\ D'Odorico, M.\ Reuter, C.\ Ripken, and R.\ Toriumi for helpful discussions and W.\ Houthoff and A.\ Kurov for constructive feedback on the manuscript.  The research of F.~S.
is supported by the Netherlands Organisation for Scientific
Research (NWO) within the Foundation for Fundamental Research on Matter (FOM) grants 13PR3137 and 13VP12.

%--------------------------------------------------------------------

%--------------------------------------------------------------------

\end{document}